# Moore's Law in CLEAR Light

Shuai Sun[1], Vikram K. Narayana[1], Tarek El-Ghazawi[1], and Volker J. Sorger[1]

[1]Department of Electrical & Computer Engineering, George Washington University, Washington DC, 20052, USA
*Correspondence to: sorger@gwu.edu

**Abstract**

**The inability of Moore's Law and other figure-of-merits (FOMs) to accurately explain the technology development of the semiconductor industry demands a holistic merit to guide the industry. Here we introduce a FOM termed CLEAR that accurately postdicts technology developments since the 1940's until today, and predicts photonics as a logical extension to keep-up the pace of information-handling machines. We show that CLEAR (Capability-to-Latency-Energy-Amount-Resistance) is multi-hierarchical applying to the device, interconnect, and system level. Being a holistic FOM, we show that empirical trends such as Moore's Law and the Makimoto's wave are special cases of the universal CLEAR merit. Looking ahead, photonic board- and chip-level technologies are able to continue the observed doubling rate of the CLEAR value every 12 months, while electronic technologies are unable to keep pace.**

**Main body:**

Fundamental physics, process control, and economic pressure demand ongoing changes and adaptations for technology development of the semiconductor industry [1]. Moore's law itself has shifted its driving-forces more than once before; from counting transistors that the industry pivoted to transistor size- and cost-scaling due to the limits of on-chip size and complexity [2]. A second transition occurred when the clock frequency became limiting due to the power density dissipation constraints described by Dennard Scaling [3]. With transistor scaling nearing fundamental physical limits, the count of transistor continues to increase, for now, driven by the parallelism introduced with multi-core and massively parallel heterogeneous architectures. This, however, worsens the communication bottleneck, while leading to the necessity to turn-off certain areas of the chip (i.e. 'dark silicon') [4]. As such the device growth rate adjusted from the initial doubling every 12-month to only 24-month today (Fig. 1, light green dots). In addition to Moore's law, other metrics were introduced to capture the computer performance trend over time. For instance the Makimoto's Wave observes cycles in the industry driven by technology standardization and customization and been quantified in a FOM, which predicts the democratization of computing (Eqn. 1).

$$Makimoto's\ FOM = \frac{MIPS}{(Size) \times (Cost) \times (Power)} \qquad (1)$$

This FOM includes a performance-cost model to track the evolution of the semiconductor industry. Similar to the one-factor based Moore's law considering only the number of transistors, Makimoto estimated compute-performance in units of million instructions per second (MIPS). Such performance enhancement, comes as an energy, space and economic cost overhead, forming a trade-off (Eqn. 1). This four-factor FOM traces the actual semiconductor industry evolution longer than Moore's law, yet it still deviates past the 1970s (Fig. 1, purple dots).

Here we explain why these deviations take place, and introduce a universal FOM that accurately predicts the development trend continuously since the 1940 until the present day. The quantifiers considered in previous FOMs were unable to capture the holistic driving forces required to accurately track the technology development pace. Analyzing the history of compute performance, the transistor count in Moore's law initially (1958-1965) tracks the 2×/year growth rate well (dashed light green line, Fig. 1), However, the size, power and cost for scaling of transistors (Dennard Scaling and Moore's second law) became a dominant factor during the next period (1965-1977). Therefore, Moore's law started to deviate from the 2× trend while Makimoto's FOM still maintained its original growth rate. Starting from the late 1970s, both the size and power scaling slowly become saturated due to the fabrication yield, energy leakage and heat dissipation challenges. Together with the emergence of parallelism (i.e. multi-core processors), Makimoto's FOM finally deviates as well (starting around 1978). As such, a new holistic FOM is needed to guide technology evolution and transition into the next-generation technologies smoothly. Such a FOM has to be holistic as to encompass performance parameters of a multitude of technology options while keeping track of the respective costs, and is given by

$$CLEAR = \frac{Capability}{(Latency) \times (Energy) \times (Amount) \times (Resistance)} \qquad (2)$$

The individual factors in of CLEAR in Eqn. 2 are defined based on the targeted hardware hierarchy at which it is applied to; it can be applied to the device, link, network, or system level. For instance, at the system level, the *capability* (C) is the data capacity (bits/second) that the system can handle, which can be obtained by multiplying the compute performance in (MIPS, millions of instructions per second) with the length of the instruction set; whereas for a link level CLEAR, the *capability* (C) represents the actual data rate of a noisy channel delivered across a certain distance between sender and receiver [5]. In general, *capability* describes a data handling ability while taking into account constrains such as the signal to noise ratio (SNR), channel crosstalk, and device bandwidth. The to-be-minimized *cost* for different hierarchy applications are similar consisting of 1) the *latency* (L) which relates to the clock frequency, 2) the *energy efficiency* (E) needed for the computation and delivery of data, and 3) the physical *amount* (A) representing the footprint of a 2D system (devices, links) or the volume of a 3D system (computers, computer cluster) of a certain technology option needed to provide a required performance. Lastly, 4) economic *resistance* (R) is a quantifier capturing effects of market forces such as technology improvement, labor efficiency, industrial structure, per unit cost, and is derived from experience models [6]. Applying this five-factor CLEAR FOM to the compute-performance vs. time data-set, we find that this metric tracks the empirical compute performance evolution remarkably well, including developments in recent years (blue dots, Fig. 1). Interestingly, fitting CLEAR results in a time-for-doubling performance of precisely 12-months. This suggests that the actual computing evolution growth rate is rather constant over more than 75 years, even among different technologies. This constant improvement trend constitutes a performance pressure against which emerging technologies need to perform in order to become viable and adoptable. The latter is exemplified by the transition between vacuum tubes (Fig.1, dark green dots) for early stage computers and transistors appearing disjoint with respect to the component count trend, while CLEAR reveals a steady improvement. Although there is no commercial optical-based computer with on-package silicon photonic interconnects today, the silicon photonic chip envisioned by IBM by the year 2020 suggests that photonics-augmented computing may continue CLEAR steadily beyond CMOS limitations (red dot Fig. 1) [7]. An interim conclusion is that the holistic CLEAR FOM describes technology development accurately amongst multiple technologies. Next we explore a case study for performance comparison between electronics interconnects against emerging photonic-plasmon hybrid links.

CASE STUDY: MAINSTREAM VIABILITY OF ON-CHIP INTEGRATED OPTICS

On-chip photonic interconnects have recently shown high-data capacity outperforming conventional electrical interconnects at the intra-chip level when hybridized with active plasmonic devices [8]. While optical data-routing is perceived as a possible solution to address the communication bottleneck of compute cores, integrated photonics has yet to be introduced into consumer electronics. This is surprising, since previous studies showed superior performance for photonic-plasmonic hybridization with a break-even signal propagation distance of ten's of micrometers relative to electronics [8]. Thus, the question arises as to why integrated optics has not been used in intra-chip interconnections in mass-market products to-date?

Towards answering this question we compare CLEAR for electronic links against hybrid photon-plasmonic links as a function of time evolution and signal propagation distance (Fig 2). The results show that the CLEAR surfaces of electronics and plasmonic-photonic hybrids exhibit break-even points (surfaces crossing, Fig. 2) that are strongly time and distance dependent. Interestingly, in the year 2016, this crossing has yet to reach chip scale (CS = 1 cm) lengths explaining why electronics is still being commercially used in network-on-chips to-date. Despite the hybrid photonic-plasmonic option offering a superior performance, a transistor nowadays only cost one-billionth of a photonic device price or less [9]. As technology and manufacturing processes improve, the performance-per-cost (i.e. CLEAR) break-even distance shortens, due to a flatter cost curve of electronics compared to photonics, the latter following a power law with time. Moreover, cost starts to increase upon further electrical interconnects density scaling associated with added overhead due to fundamental physical challenges at sub-10 nm transistor nodes [10]. In contrast, hybrid optical links, while currently being costly, are yet to benefit from economy-of-scale fabrication processing, which is the aim of the American Institute for Manufacturing Integrated Photonics (AIM Photonics). Such scaling is now possible since the recent advances in nanophotonics; the concept of enhancing light-matter-interaction allow for wavelength-compact optoelectronic devices with the benefit of high energy efficiency and fast operating speed due to the low electrical capacitance [11]. As a result, the break-even signal propagation distance between electronics and hybrid photon-plasmonic links is expected to further shift to shorter distances with development time (Fig. 2). Indeed, the CMOS-based silicon photonic chip demonstrated by IBM in 2015 is close to the break-even area of two technologies [12].

In addition to applying CLEAR to the interconnect-level, it can be amended to guide the development of the device- and network-level alike, provided minor modifications are applied; for the device-level, the signal distance becomes the device length and hence cancels in Eqn. (2). Thus the area in (2) reduces to the device scaling length. The data capacity and latency from the link becomes the device operating speed and response time, respectively. For a network-level deployment, the aggregate bandwidth per node becomes the data-handling performance that a network needs to provide, however, subject to the end-to-end latency, energy consumptions and chip-size, as well as the fabrication cost. Furthermore, CLEAR is a universal techno-economic metric not only due to its broad hierarchical applicability, but also owing to its ability to be configurable for a specific application. This can be achieved by adding weighting exponents to each factor in Eqn. (2); for instance if a system is sensitive to the power budget, the energy exponent can be increased, i.e. CLEAR becomes $CLE^2AR$, or, if a reduced footprint is critical $CLEA^2R$ may apply. Furthermore, we can perceive future links or networks to be dynamically reconfigured enabling the chip, or subsections thereof, to change its ideal operating point to shift depending on the current application, load, battery state, etc. As such CLEAR is not only a tool for road-mapping efforts and outlook forecasting, but also path towards a hardware-enabled 'smart' computer control platforms, where performance-cost tradeoffs are reassessed and optimized in real time. As such CLEAR can be regarded as the new Moore's Law that holistically captures technology development trends of a variety of hierarchical application levels which is detailed discussed in one of our work [13].

We conclude that CLEAR is a universal metric, which includes all-relevant physical and economical related factors known to-date to benchmark both electronic and optical technology options while spanning multiple hierarchical levels in computing and data communication. Based on techno-economic modeling, we show that CLEAR enables smart computer systems via application-driven dynamic reconfigurability. Founded on fundamental physics principles, it can be regarded as the next Moore's law for the coming decades in data processing and computing in order to reveal the actual technology evolution.

**Funding**

This work was supported by the Air Force Office of Scientific Research (AFOSR) under Grant FA9550-15-1-0447.

**Figures**

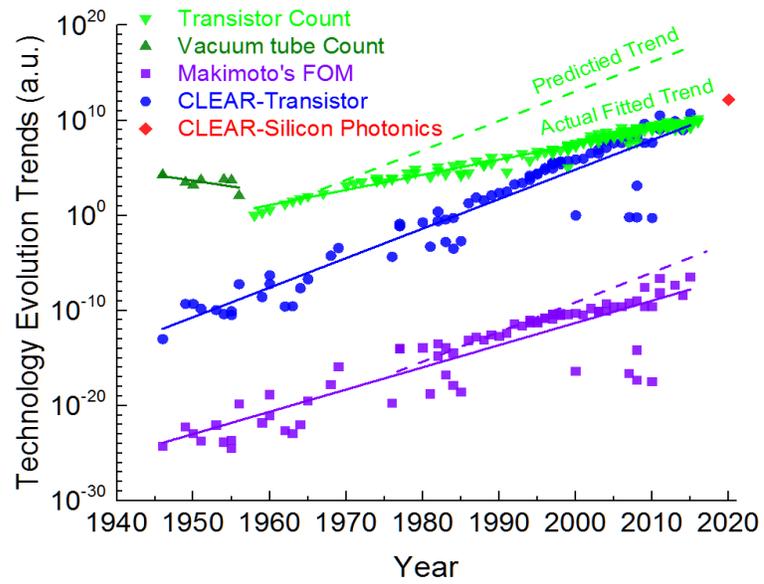

Fig. 1 The technology evolution trends over years (1946-2015). All the data are collected from various sources. Solid lines represent the linear fitting of each set of data points. Dashed lines represent the predicted 2×/year growth rate for each set of date points start from their deviation year.

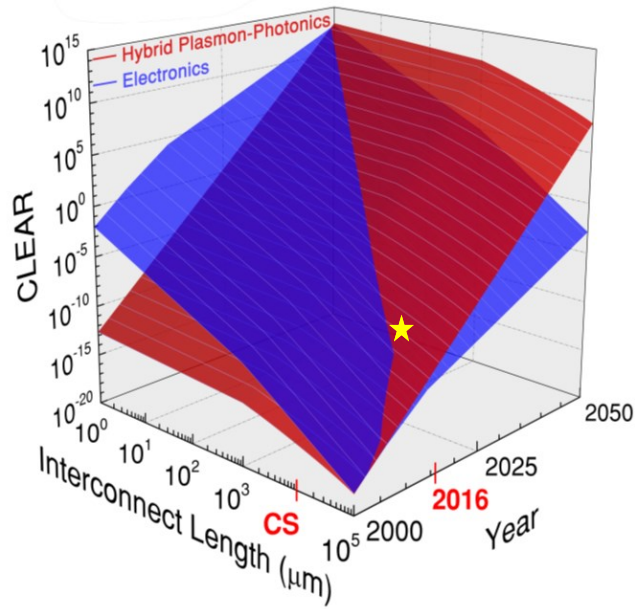

Fig. 2 Capability-to-Length-Energy-Area-Resistance (CLEAR) based performance-cost comparison between electrical (red) and hybrid photon-plasmon (blue) on-chip interconnect links as a function of link length and technology evolution time. The chip scale (CS = 1 cm) link length and current year (2016) are denoted in red. The following models are deployed; a) A capacity-area model based on the number of transistors and on-chip optical devices, which can be regarded as the original Moore's Law model; b) An energy efficiency model is derived based on Koomey's law, which is bounded by the $k_B T \ln(2) \approx 2.75$ zJ/bit, Landauer limit ($k_B$ is the Boltzmann constant; $T$ is the temperature); c) A the economic resistance model based on technology-experience models and at the year 2016, the electronic link cost less than one billionth to one millionth of the cost of the hybrid link; and d) A model for parallelism (after year 2006) capturing multi-core architecture and the limitation from 'dark' silicon concepts in electrical link interconnects. The yellow data point represents the actual CMOS silicon photonic chip that IBM fabricated in 2015.